\documentclass[aps,prb,twocolumn,groupedaddress,showpacs,amssymb,amsmath,letterpaper]{revtex4}
\usepackage{graphicx}
\usepackage[final,dvips]{epsfig}
\usepackage{color}
\usepackage{dcolumn}
\usepackage{bm}
\usepackage{array}
\usepackage{amssymb}

\definecolor{Green}{rgb}{0.2,0.96,0.2}
\definecolor{Remarks}{rgb}{1,0.3,0.3}
\definecolor{Extra}{rgb}{0.2,0.2,1}
\definecolor{Blue}{rgb}{0.2,0.3,1}
\definecolor{Black}{rgb}{0,0,0}
\definecolor{CC}{rgb}{0.2,0.75,0.2}

\newcommand{\bi}{\begin{itemize}}
\newcommand{\ei}{\end{itemize}}
\newcommand{\be}{\begin{equation}}
\newcommand{\ee}{\end{equation}}
\newcommand{\ba}{\begin{eqnarray}}
\newcommand{\ea}{\end{eqnarray}}

\newcommand{\upa}{\uparrow}
\newcommand{\dna}{\downarrow}

\newcommand{\COMMENTED}[1]{}

\newcommand{\ket}[1]{|#1\rangle}

\newcommand{\ob}[1]{\langle #1\rangle}

\newcommand{\bfp}{\mathbf{p}}

\newcommand{\bfr}{\mathbf{r}}

\newcommand{\calH}{{\cal H}}

\begin{document}

\title[Spin- and charge-density waves in the Hartree-Fock ground state...]{Spin- and charge-density waves in the Hartree-Fock ground state of the two-dimensional Hubbard model}

\author{Jie Xu, Chia-Chen Chang\footnote{Current address: Department of Physics, University of California, Davis, CA 95616, USA}
 , Eric J. Walter, Shiwei Zhang}
\affiliation{Department of Physics, College of William and Mary, Williamsburg, VA 23187, USA}

\begin{abstract}
The ground states of the two-dimensional
repulsive Hubbard model
are studied within the unrestricted Hartree-Fock (UHF) theory.
Magnetic and charge properties are determined by systematic, large-scale, exact numerical
calculations, and quantified as a function of electron doping $h$.
In the solution of the self-consistent UHF equations,
multiple initial configurations and simulated annealing are used
to facilitate convergence to the global minimum.
New approaches are employed to minimize finite-size effects in order to
reach the thermodynamic limit. At low to moderate interacting strengths
and low doping,
the UHF ground state is a linear spin-density wave (l-SDW),
with antiferromagnetic order and a
modulating wave.
The wavelength of the modulating wave is $2/h$.
Corresponding charge order exists but
is substantially weaker than the spin order, hence holes are mobile.
As the interaction is increased, the l-SDW states evolves into several different phases,
with the holes eventually becoming localized. 
A simple pairing model is presented with analytic calculations for low interaction strength and
small doping,
to help understand the numerical results and
provide a physical picture for the properties of the SDW ground
state.
By comparison with
recent many-body calculations, it is shown that, for intermediate interactions, the UHF solution 
provides a good description of the
magnetic correlations in the true ground state of the Hubbard model.
\end{abstract}


\maketitle

\section{Introduction}
\label{sec:intro}

The Hubbard model is one of the most fundamental models in quantum
physics. Despite numerous analytic and numerical investigations
\cite{Penn1966, Zitzler2002, Dagotto1994, Scalapino1994, MontorsiHubBook, RasettiHubBook},
key questions still remain about the
properties of this model \cite{Maier2005, Senechal2005, Capone2006, Zhang1997, Aimi2007}.
Surprisingly, even at the mean-field level,
its phase diagram has not yet been 
fully determined, and the ground state magnetic properties are not 
completely known.

The Hubbard model was originally proposed to describe correlations
between $d$-electrons in transition metals \cite{Hubbard}.  At
half-filling (one electron per lattice site), it gives a simple
description of the so-called Mott insulator, with antiferromagnetic
(AFM) order.  Soon after the discovery of high-$T_{\rm c}$ cuprate
superconductors, it was pointed out that the two-dimensional (2D)
Hubbard model might be an appropriate minimal model for high-$T_{\rm c}$
cuprates \cite{Anderson1987}, because of the copper-oxygen plane
geometry and the proximity of the superconducting transition to the
AFM phase of undoped mother compounds.  The 2D Hubbard model has since
become a focal point of research in condensed matter and quantum
many-body physics.

Recently, rapid experimental progress in optical lattice emulators
\cite{Emulator} has promised a new way of to approach Hubbard-like
models.  Using ultra-cold fermionic atoms trapped in periodic
laser-field potentials, these highly controllable experiments are
capable of potentially `simulating' the Hubbard model directly.
Thus the properties of the Hubbard model are not only of importance
theoretically but can also be of direct experimental relevance.

The Hartree-Fock (HF) method is the simplest paradigm to describe a
quantum many-fermion system.  The method finds the single Slater
determinant wavefunction which minimizes the variational energy.  As
is well known, the mean-field approximation involved can turn out to
be very severe.  Nevertheless the HF method has often provided the
foundation for our qualitative understanding of many systems in
condensed matter and quantum chemistry.  For example, HF correctly
predicts an AFM order in the ground state of the Hubbard model at
half-filling, even though the strength of the AFM order is
overestimated and translational symmetry is (necessarily) broken. In
quantum chemistry, HF is the starting point for most calculations and
serves as the basis for understanding the electronic structure of many
systems.

Because correlation effects (e.g., the correlation energy, which is a
fundamental concept in the framework of density functional theory)
\cite{MartinDFTBook} are often defined using the HF solution as a
reference, qualitative and quantitative understanding of the HF state
is of key importance. This has not always been easy to achieve.  For
example, the nature of the unrestricted HF (UHF) state in the electron
gas at high and intermediate densities was only recently determined
\cite{ZhangHF2008}.

In the Hubbard model, HF calculations are in principle
straightforward.  The 2D Hubbard model has been studied within the
HF approximation in some of the pioneering works on high-$T_{\rm c}$
superconductors. Inhomogeneous states have been found at small
dopings, such as spin polarons \cite{Su1988}, domain walls
\cite{Zaanen1989, Schulz1990, Su1991}, and spin density waves
\cite{Machida1989, Littlewood1991, Ichimura1992}(SDW),
and phase diagrams have been proposed
\cite{Penn1966, Ichimura1992, Igoshev2010}. 
Due to computing power limitations, however,
these studies have either done exact numerical calculations
at only a few doping and interaction parameters \cite{Schulz1990, Su1991, Littlewood1991},
or have scanned parameters with restricted forms of the solution \cite{Ichimura1992, Schulz1990,Igoshev2010}.
Furthermore, finite-size effects were difficult to remove,
as we discuss below, which can mask the true solution in the
thermodynamic limit.
A systematic and quantitative understanding of the magnetic properties of the 
UHF ground state has not been achieved.

In this work, we perform extensive numerical calculations to
determine the exact UHF ground state of the Hubbard
model in the low to intermediate interacting strength regime.
The exact UHF ground state we achieved is a full numerical solution
of HF Hubbard Hamiltonian
(see Sec.~\ref{sec:method} Eqs.~(\ref{HFHamiltonian}) and (\ref{CoupledEq})),
as opposed to constrained searches or non-self-consistent solutions.
We study the
spin and charge properties as a function of
interacting strength and doping concentration. Full numerical
solutions of the UHF equations are computed using twist-averaged
boundary conditions for system sizes well beyond those previously studied.
We also present a simple pairing model, with analytic calculations at low doping and small
interacting strengths, to complement the numerical results and
provide a qualitative physical picture of the magnetic properties of the model.

Our combined numerical and analytical calculations show that, at a
finite doping $h$, the UHF ground state at low
and intermediate strengths $U/t$ is a static linear SDW (l-SDW)
state.  As the interaction strength is raised beyond a
critical value, l-SDW order develops along the $[10]$-direction,
accompanied by a weaker linear charge density wave (l-CDW).  The
characteristic wavelength of the l-SDW is found to be $2/h$ and the
wavelength of the corresponding l-CDW is $1/h$.  As the interaction
strength is increased, stripe or domain walls states develop along the
diagonal $[11]$-direction, in which the holes are localized. The
diagonal stripe (d-stripes) state and the l-SDW state are separated by
either a linear stripe state (l-stripes, along $[10]$-direction) or a
diagonal SDW (d-SDW) state, depending on the doping.  These are
summarized with a UHF phase diagram for interaction up to $U/t\sim 10$
and doping up to $h \sim 35\%$

The remainder of the paper is organized as follows.  In
Sec.~\ref{sec:method}, the self-consistent scheme used for solving the
mean-field Hubbard model is summarized. The numerical results are
presented in Sec.~\ref{sec:result}, and analytic calculations are
described in Sec.~\ref{sec:model}.  In Sec.~\ref{sec:discussion} the
results are discussed and summarized in a phase diagram, and we
conclude the paper in Sec.~\ref{sec:conclusion}.

\section{Method}
\label{sec:method}

The Hamiltonian of the single-band repulsive Hubbard model reads
\begin{equation}
  \calH = -t \sum_{\{\bfr\bfr'\},\sigma}
               \left( c_{\bfr \sigma}^\dagger c_{\bfr' \sigma} +
                        c_{\bfr' \sigma}^\dagger c_{\bfr \sigma} \right)
               + U\sum_\bfr n_{\bfr \uparrow} n_{\bfr \downarrow},
\label{Hamiltonian}
\end{equation}
where $U>0$ is the interacting strength and $t$ is the hopping amplitude
between nearest neighbor sites (denoted by $\{\bfr\bfr'\}$ in the summation).
Throughout this work, energy is quoted in units of $t$ and we set $t=1$.
The operator $c_{\bfr\sigma}^\dagger$ ($c_{\bfr\sigma}$) creates (annihilates)
an electron with spin $\sigma$ ($\sigma=\uparrow, \downarrow$)
at site index $\bfr$, which runs through the lattice of
size $N=L_x \times L_y$.
The total number of spin-$\sigma$ electrons is denoted by $N_\sigma$,
and we assume that the system has no spin polarization,
i.e. $N_\uparrow=N_\downarrow$.
Under this assumption, the model has only two parameters,
namely, the onsite repulsion $U$ and the doping
\begin{equation}
h\equiv
1-(N_\uparrow+N_\downarrow)/N\equiv N_{\rm hole}/N,
\label{eq:def_doping}
\end{equation}
where we have used $N_{\rm hole}$ to denote the number of holes in the system.
Due to particle-hole symmetry, we confine ourselves in the region where
$N_\upa + N_\dna \leq N$. Therefore the total density is given by $\ob{n}=1-h$.

Standard linearization of Eq.~(\ref{Hamiltonian}) leads to the
mean-field HF Hamiltonian:
\begin{equation}
  \calH_{\rm HF} = \calH_{\rm HF}^\uparrow + \calH_{\rm HF}^\downarrow,
\label{HFHamiltonian}
\end{equation}
with
\ba
  \calH_{\rm HF}^\sigma &=&
    -t \sum_{\{\bfr\bfr'\}}
      \left( c_{\bfr \sigma}^\dagger c_{\bfr' \sigma} +
             c_{\bfr'\sigma}^\dagger c_{\bfr\sigma} \right) \nonumber \\
     && + U \sum_\bfr n_{\bfr\sigma} \langle n_{\bfr\bar{\sigma}} \rangle
     -\frac{1}{2}U \sum_\bfr \langle n_{\bfr\uparrow}\rangle \langle n_{\bfr\downarrow}\rangle,
\label{CoupledEq}
\ea
where $\bar\sigma$ is the conjugate of $\sigma$ and
$\langle n_{\bfr\bar{\sigma}}\rangle$ is an average density.
The mean-field decoupling employed in Eq.~(\ref{HFHamiltonian}) assumes
the $z$-axis as the quantization direction, thus breaking the spin rotational
symmetry of the Hubbard Hamiltonian (\ref{Hamiltonian}).
After fixing the quantization orientation and requiring no spin polarization,
the solution of the HF Hamiltonian is restricted to the $S^z=0$ sector,
i.e. spin textures in the $x$-$y$ plan, for instance spiral SDWs, are excluded.
 (At low $U$, the solutions turn out to be l-SDWs. Then a single spiral
 cannot be the ground state, since a left-handed spiral can always be
 combined with a right-handed one, or vice versa,
 to make an l-SDW which has lower energy. \cite{ZhangHF2008})

For a given set of parameters $(U,N,N_\uparrow,N_\downarrow)$,
the HF Hamiltonian (\ref{HFHamiltonian}) is numerically solved using
a self-consistent scheme.
We begin the procedure by selecting a trial solution
in the form of a single Slater determinant
for each spin component:
\begin{equation}
 \Phi^{(0)}_\sigma =
\left(
 \begin{array}{cccc}
  \phi_\sigma^{11} & \phi_\sigma^{12} & \cdots & \phi_\sigma^{1N_\sigma}\\
  \phi_\sigma^{21} & \phi_\sigma^{22} & \cdots & \phi_\sigma^{2N_\sigma}\\
  \vdots    & \vdots    & \ddots & \vdots    \\
  \phi_\sigma^{N1} & \phi_\sigma^{N2} & \cdots & \phi_\sigma^{NN_\sigma}
 \end{array}
\right ),
\label{WaveFunc}
\end{equation}
where each column is normalized to $1$.
In the restricted HF (RHF) method, spin-$\upa$ and spin-$\dna$
parts of the total wavefunction are the same:
 $\Phi^{(0)}_\uparrow=\Phi^{(0)}_\downarrow$.
The RHF method always gives the non-interacting solution
 in the systems studied in this paper.
In the UHF method, which is adopted here,
$\Phi^{(0)}_\uparrow$ and $\Phi^{(0)}_\downarrow$ are allowed to differ
and they converge via the coupled Eqs.~(\ref{CoupledEq}).
The trial densities at site $\bfr$ can be expressed as
\begin{equation}
  \langle n^{(0)}_{\bfr\sigma} \rangle =
  \left[\Phi^{(0)}_\sigma \left(\Phi^{(0)}_\sigma\right)^H\right]_{\bfr\bfr},
\label{DenFormula}
\end{equation}
where `$H$' indicates conjugate transpose of the matrix, and
we have assumed that the orbitals in $\Phi^{(0)}_\sigma$ are orthonormal.
An
$N\times N$ matrix $M^\uparrow$ ($M^\downarrow$)
for $\calH_{HF}^\uparrow$ ($\calH_{HF}^\downarrow$) is then constructed from the
densities.
By exactly diagonalizing $M^\sigma$, we obtain the energy
\begin{equation}
E^{(1)}_\sigma = \sum_{i=1}^{N_\sigma}\,\lambda^{(1)}_{\sigma i},
\end{equation}
where $\lambda^{(1)}_{\sigma 1} < \lambda^{(1)}_{\sigma 2} < \lambda^{(1)}_{\sigma 3} < \ldots < \lambda^{(1)}_{\sigma N_\sigma}$
are the lowest $N_\sigma$ eigenvalues of $M^\sigma$.
The wavefunction $\Phi^{(1)}_\sigma$ is obtained by filling up $N_\sigma$
corresponding orbitals of $\lambda^{(1)}_{\sigma i}$.
The new density $\langle n^{(1)}_{\bfr\upa} \rangle$ ($\langle n^{(1)}_{\bfr\dna} \rangle$)
is then calculated from $\Phi^{(1)}_\uparrow$ ($\Phi^{(1)}_\downarrow$),
which is used to update $M^\downarrow$ ($M^\uparrow$).
We iterate this process until the total
energy $E^{(\ell)}=E^{(\ell)}_\uparrow+E^{(\ell)}_\downarrow$
and the density $\langle n^{(\ell)}_{\bfr\sigma} \rangle$ is converged.

Care must be taken when updating the density during the iteration.
As is typical in self-consistent algorithms, convergence to a fixed point
is not guaranteed if $\ob{n_{\bfr\sigma}^{(\ell-1)}}$ is taken directly as
an input for the $\ell$-th step.
To improve convergence, we adopt a mixing scheme:
The $\ell$-th input density is constructed as a linear
combination of previous input and output densities as:
\begin{equation}
  \langle n^{(\ell),{\rm in}}_{\bfr\sigma} \rangle
  = (1-\alpha) \langle n^{(\ell-1),{\rm in}}_{\bfr\sigma} \rangle
  +\alpha \langle n^{(\ell-1),{\rm out}}_{\bfr\sigma} \rangle,
\label{MixDensity}
\end{equation}
where `in' indicates the input density to construct $M^\sigma$,
and `out' denotes the output density calculated by diagonalizing $M^\sigma$.
The mixing parameter $\alpha$ is typically chosen to be between $\sim 0.5$ and $0.75$.

Due to non-linearity of the coupled Eqs.~(\ref{CoupledEq}),
we implement two additional procedures to help the system
reach the global minimum.
Firstly, different initial wavefunctions are used and the
consistency between the results is checked.
Secondly, we perform multiple annealing cycles:
in each cycle a random perturbation (whose strength can
be controlled) is applied to the converged solution
and the self-consistent process is repeated.

To reduce shell and one-body finite-size effects,
we use twist-averaged boundary conditions
(TABC)\cite{Poilblanc1991,Gross,Lin2001}, under which
the wavefunction $\Psi(\mathbf{r}_1,\mathbf{r}_2,\ldots)$ gains a phase when
electrons hop around lattice boundaries
\begin{equation}
  \Psi(\ldots,\mathbf{r}_j+\mathbf{L},\ldots) = e^{i\widehat{\mathbf{L}}\cdot\mathbf{\Theta}}\Psi(\ldots,\mathbf{r}_j,\ldots),
\label{TBC}
\end{equation}
where $\widehat{\mathbf{L}}$ is the unit vector along $\mathbf{L}$,
and the twist angle $\mathbf{\Theta}=(\theta_x,\theta_y)$ is
an input parameter which is randomly chosen in this work.
For a given $\mathbf{\Theta}$,
the TABC is the same as a random shift of the momentum space grid.
This reduces the discretization error in the integration.
In the HF solution, the TABC is applied to each orbital, i.e., each
column in Eq.~(\ref{WaveFunc}).
With a generic $\mathbf{\Theta}$, there will be no degeneracy in the
one-electron energy levels.
We often average the results over many
random twist angles \cite{Lin2001} in each system
to improve convergence to the thermodynamic limit.
As can be seen from the energy results in Sec.~\ref{sec:result}, this procedure produces
a smooth curve vs. doping, where the one-body finite-size effect is minimized.
Additional finite-size errors, which result from the interaction and the formation of long
wavelength collective modes \cite{Chang2010}, are not removed
from this approach. We use rectangular lattices in our simulations to help
detect the l-SDW states with long modulating wavelengths, as discussed in Sec.~\ref{sec:result}.

\section{Numerical Results}
\label{sec:result}

\begin{figure}
\includegraphics[width=\columnwidth]{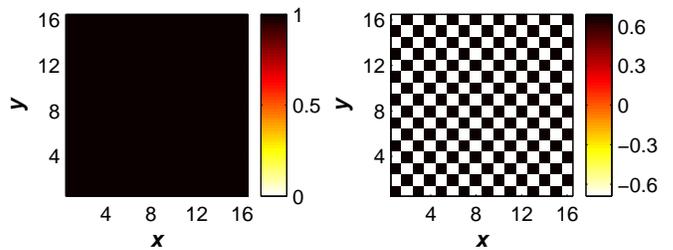}
\caption{
(Color online)
Contour plots of CD (left) and SD (right) at half-filling
and $U=4.0$ on a $16 \times 16$ lattice.
CD is uniformly distributed at a density of 1.
SD is AFM with uniform amplitude.
}
\label{halffilling}
\end{figure}

Various observables are computed with the converged UHF wavefunction.
Two quantities examined throughout the paper are the charge-density (CD) $\rho(\mathbf r)$ and the spin-density (SD) $s(\mathbf r)$
defined as
\begin{eqnarray}
  \rho(\mathbf r) &\equiv& \ob{ n_{\bfr\upa} }+\ob{ n_{\bfr\dna} }, \\
       s(\mathbf r) &\equiv& \ob{ n_{\bfr\upa} }-\ob{ n_{\bfr\dna} }.
\label{CDSDDef}
\end{eqnarray}
We will also study the converged UHF eigenvalues $\lambda_{\sigma \mathbf p}$ and
momentum distribution
$n_{\mathbf p \sigma}=\langle c_{{\mathbf p}\sigma }^\dagger c_{{\mathbf p}\sigma}\rangle$,
where $c_{{\mathbf p}\sigma}$ is, as usual, defined
by the Fourier transform of $c_{{\mathbf r}\sigma}$.
Figure~\ref{halffilling} illustrates the behavior of
the reference system at $h=0$, which is an AFM state with constant
$\rho(\mathbf r)=1$.

\begin{figure}
\includegraphics[width=\columnwidth]{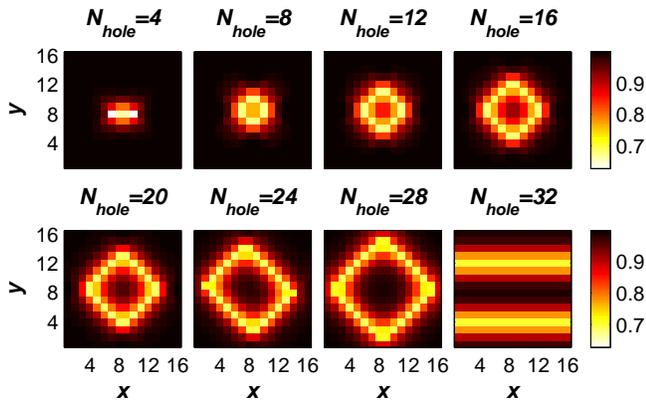}
\caption{
(Color online)
Contour plots of CD for a supercell of $16 \times 16$ at $U=4.0$
as $N_{hole}$ is increased.
The finite-size effect is strong until the l-CDW wavelength is decreased sufficiently to fit into the
simulation cell,
as shown in the right bottom plot.
}
\label{HoleGrow}
\end{figure}

In Fig.~\ref{HoleGrow}, CD in the $16\times 16$ reference system
is plotted as holes are doped into the lattice.
As doping $h$ is varied,
holes tend to cluster and form different patterns.
These patterns have a strong $h$-dependence, which is a result of
strong finite-size effects.
Here the system is at an intermediate interaction strength of $U=4.0$.
As the interaction becomes weaker, we find that the variations
in the patterns become larger and depend sensitively on $\mathbf\Theta$ (not shown).
This is similar to
what is seen in the UHF solution of an electron gas \cite{ZhangHF2008}.

\subsection{Linear spin-density wave (l-SDW) state}
\label{ssec:l-sdw}

\begin{figure}
\includegraphics[width=\columnwidth]{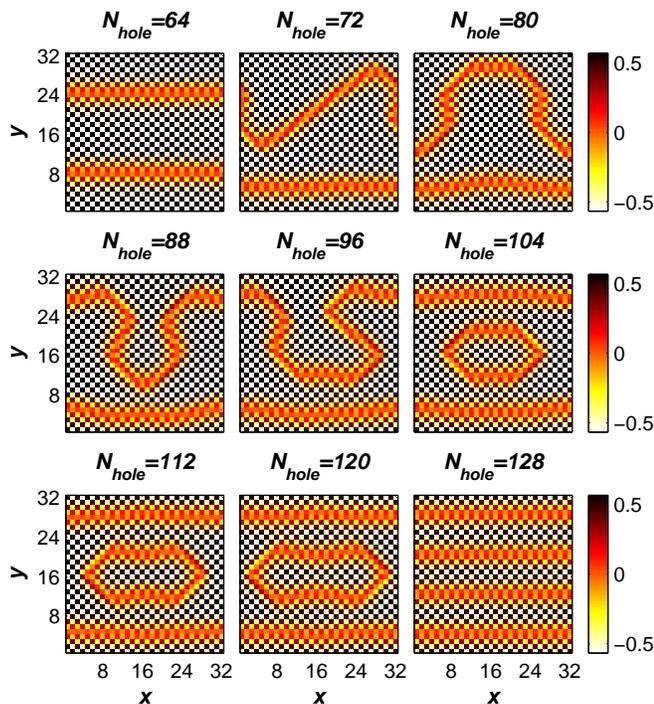}
\caption{
(Color online)
Contour plots of SD for a supercell of $32 \times 32$
at $U=3.0$ as $N_{hole}$ is increased.
An l-SDW exists when the density is such
that the supercell size is sufficient to accommodate the l-SDW,
as shown in the left top or the right bottom plot.
}
\label{32x32vary_h}
\end{figure}

We first focus on low to moderate
interacting strengths ($U <$ half of the bandwidth) and small doping ($h \lesssim 0.1$), and
examine the properties of the UHF solution as a function of doping $h$,
i.e., as the system moves away from
half-filling ($h=0$). We will show that  the UHF ground state at low and moderate $U$ is a linear
spin-density wave (l-SDW) along the [01] direction.
Figure~\ref{32x32vary_h} shows the
results
from a $32 \times 32$ supercell.
An l-SDW is seen
whenever the density is such that an l-SDW can be accommodated in
the supercell.
(The choice between $x$- and $y$-directions in the broken-symmetry
UHF state is of course random.
To help visualization in the figures, we have selected the
same direction, either by an initial bias or by rotating the final result.)
At incommensurate densities,
strong finite-size effects are present, where
the pattern of the cluster is not scalable to the thermodynamic limit.
An example is seen by comparing
$N_{\rm hole}=16$ in Fig.~\ref{HoleGrow} (not long enough for one period of SDW) and
$N_{\rm hole}=64$ in Fig.~\ref{32x32vary_h}: in both cases $h=1/16$.
$N_{\rm hole}=24$ in Fig.~\ref{HoleGrow} vs. $N_{\rm hole}=96$ in Fig.~\ref{32x32vary_h} is another
(both have $h=3/32$).
The finite-size effects will be further discussed below.

Although significantly larger lattice sizes are reached,
the pattern variation clearly indicates that care
must be taken in a numerical calculation, and additional ingredients are
needed, in order to better approach the thermodynamic limit.
We use two additional ingredients in our numerical simulations:
TABC
and rectangular supercells.
To reduce the one-body finite-size effects,
most of our results are averaged over $\sim 20$ random
$\mathbf{\Theta}$ values.
In plots showing $\mathbf{\Theta}$-averaged results,
 the statistical
uncertainties from the twist angles
are indicated by the error bars.
The residual (two-body) finite-size effects are reduced by the use of
rectangular supercells. This allows us to study longer wavelength modes
without increasing the computational cost (compared to a square lattice of the same
number of lattice sites, $N$.) Obviously rectangular lattices break the symmetry between
$x$- and $y$-directions, and can introduce an additional bias. To minimize the effect, we carry
out calculations with different supercells with varying aspect ratios to check consistency in the results.

\begin{figure}
\includegraphics[width=\columnwidth]{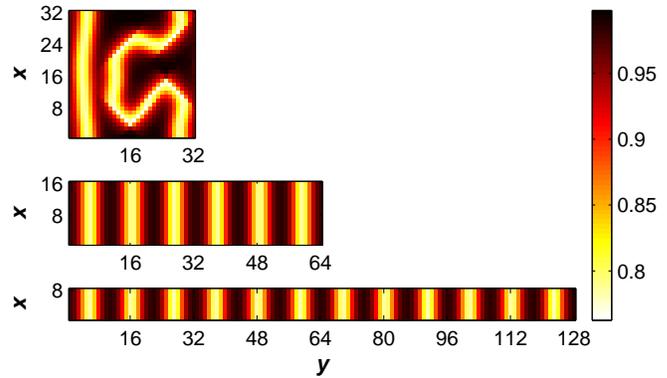}
\caption{
(Color online)
Contour plots of CD for systems of {\em fixed\/}
$N=L_x \times L_y=32 \times 32=16 \times 64=8 \times 128=1024$
(from top to bottom) and fixed doping $h=3/32$ ($N_{hole}=96$) at $U=3.0$.
A stable l-SDW solution emerges when the supercell is commensurate.
Note that only the accompanying CDW is shown here.
}
\label{fix_h_vary_LxLy}
\end{figure}

An illustrative set of results is shown in Fig.~\ref{fix_h_vary_LxLy}. We
adjust $L_x$ and $L_y$ while keeping the size
$N=L_x\times L_y=32\times 32$ fixed.
An l-SDW solution is seen
in a rectangular supercell whenever
$L_y$ is sufficiently large to accommodate a wave.
Note that the rectangular supercell does not bias the SDW in the $y$-direction
(when $L_y > L_x$). An l-SDW is observed along the $x$-direction if $L_x$ is commensurate
with the SDW wavelength. (An example of this is in Fig.~\ref{fig:energy_fixedN_varyLxLy} below,
where the solution in the $20\times 36$ lattice is two waves propagating along $x$-direction.)

\begin{figure}
\includegraphics[width=\columnwidth]{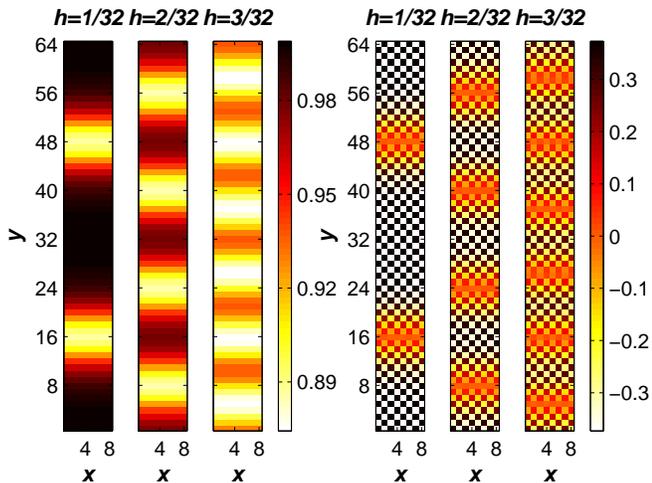}
\caption{
(Color online)
Contour plots of CD (left) and SD (right) vs. doping.
The system is an $8 \times 64$ supercell
at $U=2.0$ with doping of $h=1/32$, $2/32$ and $3/32$ (from left to right).
The wavelength of the l-CDW is $\lambda_{\rm l-CDW}=1/h$
and that of the l-SDW is $\lambda_{\rm l-SDW}=2/h$.
}
\label{CDSD_h_2D}
\end{figure}

\begin{figure}
\includegraphics[width=\columnwidth]{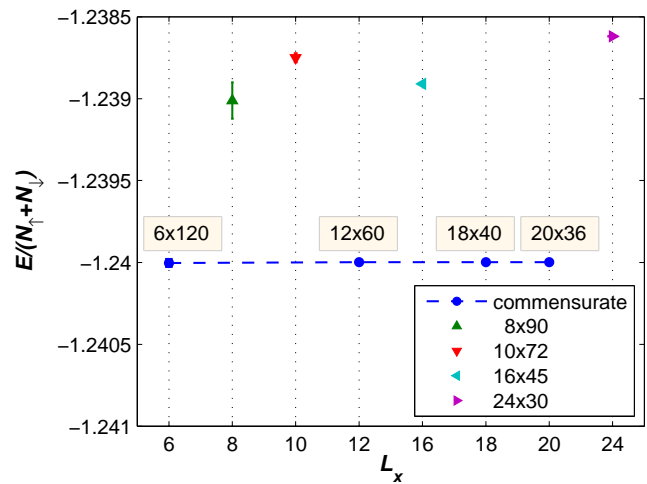}
\caption{
(Color online)
Ground state energy per particle as a function of the aspect ratio
for a series of supercells with a fixed
$N=L_x \times L_y = 720$.
Doping is at  $h=0.1$ and the interaction strength is $U=2.5$.
Results are averaged over 22 random $\mathbf{\Theta}$ values;
statistical error bars are shown, although some are too small to be seen.
For the supercells which can accommodate full l-SDW/CDW,
whose wavelengths are determined by
Eqs.~(\ref{WaveLength2D_CDW}), (\ref{WaveLength2D_SDW}),
the variational energy is consistent and lower.
}
\label{fig:energy_fixedN_varyLxLy}
\end{figure}

 From the results in Figs.~\ref{HoleGrow} and \ref{32x32vary_h},
it is clear that the wavelengths of the l-SDW and l-CDW vary with doping $h$.
The results of an $8 \times 64$ lattice with various values of $h$ are shown in
Fig.~\ref{CDSD_h_2D}.  As can be seen,
the wavelength of the l-CDW/SDW decreases with
$h$. Unlike in Fig.~\ref{32x32vary_h},
the lattice size in this case has been chosen so that
$L_y$ is commensurate with the wavelength
in each figure.  For example, there are exactly two CD waves at
$h=1/32$,
giving a wavelength of
$L_y/2=32(=1/h)$.
The wavelengths of SDW (right panel) are twice
those of CDW.  When the doping is doubled
or tripled,
the number of waves being accommodated changes accordingly,
i.e. the wavelength shortens by $1/2$ or $1/3$, respectively.
The modulating wavelengths of the
l-CDW and l-SDW are thus given by
\begin{eqnarray}
  \lambda_{\rm l-CDW}(h) &=& {1 \over h}, \label{WaveLength2D_CDW} \\
  \lambda_{\rm l-SDW}(h) &=& {2 \over h}.\label{WaveLength2D_SDW}
\end{eqnarray}
The wavelength relations are verified with many different choices
of the aspect ratio.

The variational energy
of the UHF ground state is examined
in Fig.~\ref{fig:energy_fixedN_varyLxLy}.
A series of supercells are studied with a fixed $N=L_x \times L_y=720$
and $h=0.1$, while varying $L_x$ and $L_y$.
It is seen that, for all
supercell choices commensurate with the predicted wavelength, the
energies are consistent and are lower. In systems which are
incommensurate and cannot accommodate the l-SDW/CDW, the resulting
ground state energies from the UHF solution are higher, indicating
the frustration effect in the variational solution because of the finite size of
the supercells.
In Sec.~\ref{sec:model},
we will present an analysis showing why in general the l-SDW
is favored at low $U$.

\begin{figure}
\includegraphics[width=\columnwidth]{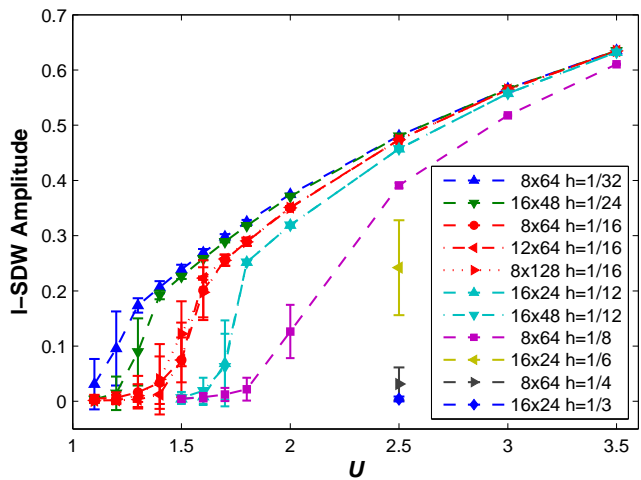}
\caption{
(Color online)
l-SDW amplitude as a function of $U$
at various doping for several supercells.
At each data point,
the result is averaged over 22 random $\mathbf{\Theta}$ values
and the error bar is the statistical error.
From left to right, the doping is increased.
At a fixed doping, different supercells give consistent results.
The amplitude increases with $U$ beyond $U_{\rm c}$ and converges to a
stripe or domain walls state (see Sec.~\ref{sec:discussion}).
}
\label{CDSD_Uh_2D}
\end{figure}

More lattice sizes at various dopings and interacting strengths are studied.
The amplitudes of the l-SDW in the
obtained solution are summarized in
Fig.~\ref{CDSD_Uh_2D}.
It can be seen that at each fixed density,
the l-SDW amplitude decreases as $U$ is decreased and eventually
vanishes, indicating the disappearance of the broken-symmetry UHF solution
at a critical interaction strength $U_{\rm c}$. Below $U_{\rm c}$ only a RHF solution
exists.
The critical value $U_{\rm c}$ appears to decrease with $h$ and approaches $0$
at zero doping. This is consistent with the situation at half-filling ($h=0$),
where the Fermi
surface (FS) is an open shell and a UHF state can be formed by `pairing' \cite{ZhangHF2008} across it with no cost to
the kinetic energy.
For a fixed $U$, the amplitude of the l-SDW decreases with doping (as does the wavelength).

The amplitude fluctuation is the strongest near $U_{\rm c}$, indicated by large statistical errors,
and decreases as $U$ is increased.
This can be understood from the mechanism for the l-SDW states in the UHF solution.
 The l-SDW state is formed by `pairing' or nesting of electrons near the FS \cite{Overhauser} (see also Sec.~\ref{sec:model}).
At low $U$, the UHF solution only contains
a small number of excitations \cite{Overhauser,ZhangHF2008} to
plane-wave states immediately beyond the FS.  In a finite-sized
system, how well the desired pairing can be achieved depends sensitively on
the particular topology of momentum space grid, and the results therefore
show more fluctuation with respect to $N$ or $\mathbf{\Theta}$.
Thus the l-SDW
amplitude is small around $U_{\rm c}$, and sensitive to the boundary conditions, giving
relatively large statistical error bars.
At larger $U$, there are more excitations
above the FS, and the plane-wave states necessary for pairing
become available independent of $\mathbf{\Theta}$, so less fluctuation is seen.

\begin{figure}
\includegraphics[width=\columnwidth]{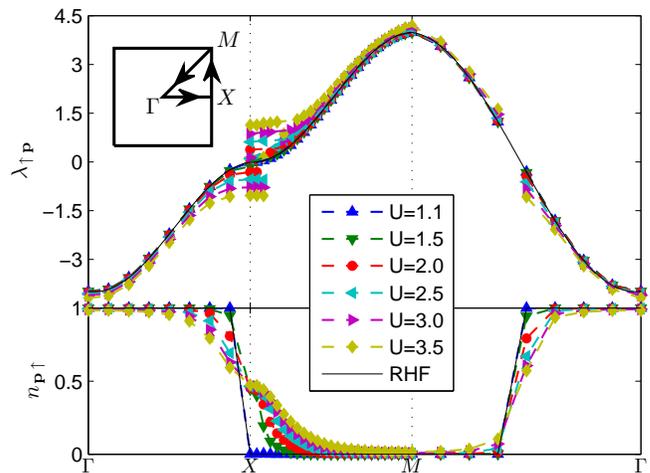}
\caption{
(Color online)
UHF eigenvalues $\lambda_{\sigma \mathbf p}$ vs. momentum ${\mathbf p}$ (top)
and corresponding momentum distribution (bottom).
Both quantities are plotted along symmetry lines in momentum space, as depicted in the inset.
The system is a $16\times 48$ supercell with doping of $h=1/24$
for a series of $U$. In the top, the RHF (non-interacting) band-structure is also shown for
comparison.
}
\label{HF_bandstructure}
\end{figure}

The picture we described above is supported by the UHF band structure
and momentum distribution shown in Fig.~\ref{HF_bandstructure}.
In the figure, we plot the UHF eigenvalues $\lambda_{\uparrow \mathbf p}$ (shifted by the
mean-field background $U\ob{n_{\mathbf r \upa}}\ob{n_{\mathbf r \dna}}/2$)
for a series of $U$.
Each $\lambda_{\uparrow \mathbf p}$ is identified
with a wavevector ${\mathbf p}$ by the maximum plane-wave component in the corresponding wavefunction, i.e., according to the magnitude of $|\langle {\mathbf p}|\phi_{\uparrow \mathbf p}\rangle|$.
The corresponding momentum distribution is also shown.
Results are the same for $\sigma=\uparrow$ and $\downarrow$ and are
only shown for spin-$\upa$  electrons.
We will omit the $\sigma$-index below unless it is necessary.
At small $U$ values ($U \lesssim 1/4$ of the bandwidth),
the deviation of $n_{\mathbf p \upa}$ from the non-interacting
(or RHF) result is not drastic.
We see that, as $U$ exceeds $U_{\rm c}$, a gap opens up in the band structure. Only
a small number of states, $|{\mathbf p}\rangle$, near the FS participate in the formation
of the broken-symmetry state. As $U$ is increased, there are more excitations and more states
becoming involved. In Sec.~\ref{ssec:l-sdw-compare-model-v-data}, we discuss the mechanism
in further detail, and show how it is described by a simple pairing model at low $U$.

As seen from Fig.~\ref{CDSD_Uh_2D},
once the interaction strength is above the immediate vicinity of $U_{\rm c}$,
the finite-size effect becomes minimal in the system sizes we have studied.
The wavelength and amplitude of the l-SDW (CDW) do not change with the supercell size.
Larger supercells give essentially identical results with the SDW replicated
to fill the (commensurate) supercell.

\subsection{Diagonal spin-density wave (d-SDW), linear and diagonal stripe (l/d-stripes) states }
\label{ssec:d-sdw-stripe}

\begin{figure}
\includegraphics[width=\columnwidth]{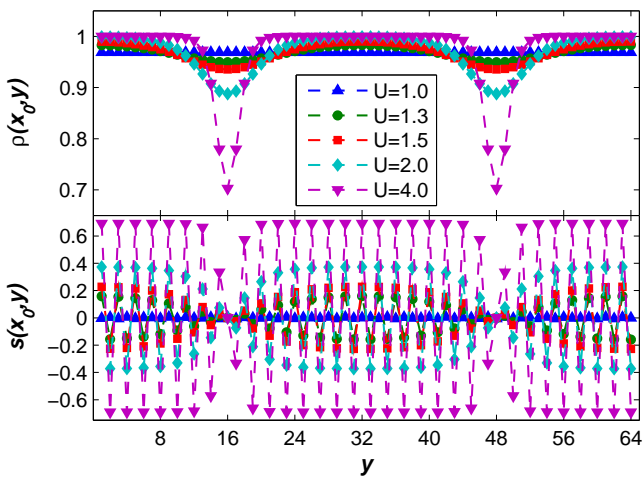}
\caption{ (Color online) CD (top) and SD
  (bottom) along $y$-direction vs. $U$. The system being studied
  is an 8x64 supercell with doping of 1/32 at $U=1.0,1.3,1.5,2.0,4.0$.
  Each curve is a 1D cut in which the linear wave propagates.
  Beyond $U_{\rm c}$, the l-CDW and l-SDW amplitudes increase with $U$
  and the ground state ends up in an l-stripes state.
  The CDW amplitude is much weaker than that of the SDW.  }
\label{CDSD_U_2D}
\end{figure}

\begin{figure}
\includegraphics[width=\columnwidth]{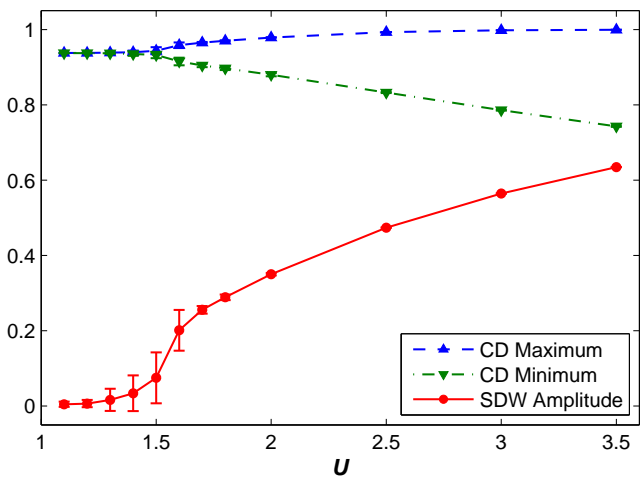}
\caption{
(Color online)
Maximum and minimum of the CD and SDW amplitude for $8 \times 64$ supercell
with doping of 1/16.
}
\label{CDmaxmin_SDamp}
\end{figure}

As the interaction strength $U$ is further increased, the UHF ground state changes character.
Figure~\ref{CDSD_U_2D} shows the CD and SD along the $y$-direction, along which the
linear wave propagates. Above $U_{\rm c}$, the amplitude of the l-SDW (and CDW) grows with
$U$.  As $U$ is further increased, the CD reaches $1$ and starts
saturating, creating deeper density valleys at the nodes of the l-SDW.
The maximum and minimum of CD and the l-SDW amplitude as a function of $U$
are plotted in Fig.~\ref{CDmaxmin_SDamp} to further illustrate this.
As discussed in Sec.~\ref{ssec:l-sdw}, CD/SD orders are developed beyond $U \sim 1.5$.
  The l-SDW amplitude is much greater
than that of the l-CDW.
The CD maximum saturates at $1$ above $U \sim 2.5$,
indicating the formation of a linear stripe (l-stripes) state.
The stripe or domain wall states differ from the SDW state because
of CD saturation, forming hole-free domains that separate regions in which the holes
are localized. The SDW state, in contrast, is a wave state in which the CD spatially oscillates
but does not reach $1$, and the holes are delocalized.

Thus at low dopings (high densities, $h \lesssim 0.1$),
 the l-SDW state
turns into an l-stripes state
as $U$ is increased, with the l-stripes along the same direction ($x$- or $y$-) and
having the same characteristic wavelength.
When $U$ is further increased the solution changes orientation, turning into a
stripes state with modulation along the $[11]$-direction, a diagonal stripes (d-stripes) state.

\begin{figure}
\includegraphics[width=\columnwidth]{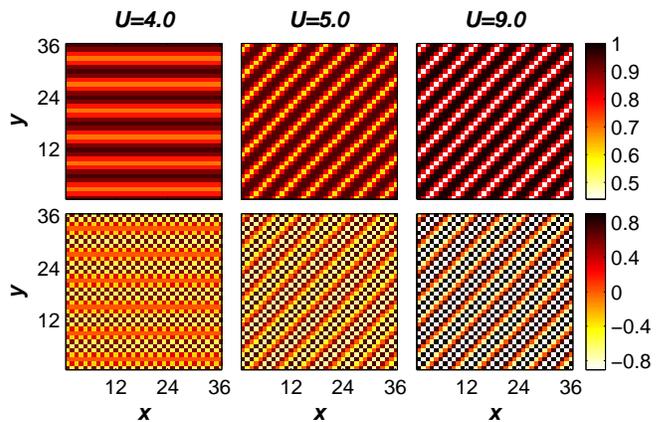}
\caption{
(Color online)
Contour plots of CD (top) and SD (bottom) vs. interacting strengths.
The system being studied is a $36 \times 36$ supercell
with doping of $h=1/6$ at $U=4.0$, $5.0$ and $9.0$ (from left to right),
representing l-SDW, d-SDW and d-stripes state respectively.
}
\label{diag_CD}
\end{figure}

At somewhat larger doping ($0.1 \lesssim h \lesssim 0.3$), the evolution of the l-SDW state with $U$ is different.
The SDW state changes its modulation direction
from the $[10]$-direction to diagonal.
 (d-SDW has been discussed in
Ref.~\onlinecite{Littlewood1991}, for example.) Figure~\ref{diag_CD} shows an example for
doping of $h=1/6$.
We see that the modulating wave changes direction
from $[10]$ at $U=4.0$ to $[11]$ at $U=5.0$,
and the d-SDW saturates to become a d-stripes state at $U=9.0$.

We have scanned different parameter combinations to map out the sequence of the
evolution of the UHF ground state.
In Sec.~\ref{sec:discussion}, a phase diagram is sketched to summarize the
properties of the UHF ground state in the part of the phase space on which we have focused.
The difference in the pairing mechanism of the d-SDW state from that of the l-SDW state is briefly discussed
in Sec.~\ref{ssec:D-sdw-model}.

\section{Analytic Calculations}
\label{sec:model}

In this section we present a phenomenological model of the l-SDW state at low $U$ and small
$h$. The model will help explain the numerical findings and
provide a simple physical picture that captures the basic features of
the exact UHF solutions in this parameter regime.  The numerical studies are
independent of the analysis here, but together they will give a more
complete description of the UHF states.
Below we first discuss the basic pairing model  \cite{Overhauser,ZhangHF2008},
then carry out calculations in detail in the limit of small $U$ and $h$ for the l-SDW state,
which is the focus of the present work. Some quantitative comparisons and validations
of the pairing
analysis are then presented, using the numerical data from calculations presented in
Sec.~\ref{ssec:l-sdw}. We then briefly discuss the mechanism for d-SDW and d-stripes orders
at higher $U$.

\subsection{Pairing model}
\label{ssec:pairing}

At low $U$, the region
of interest in momentum space is the
immediate vicinity of the FS, where pairing
effects of electrons determine the nature of the
UHF solution.
(Often the effect has been discussed in the context of nesting. 
We refer to the mechanism as pairing since, 
although nesting greatly facilitates pairing in the Hubbard model, 
it is not required for the pairing mechanism to be realized, as seen in the electron gas\cite{ZhangHF2008}.)
In the fully filled region inside the FS, the
electron density is uniform,
\begin{equation}
  n_{\bfp\sigma}(\bfp) = \bar n_\sigma = \frac {N}{4\pi^2}.
\label{den_k}
\end{equation}

\begin{figure}
\includegraphics[width=\columnwidth]{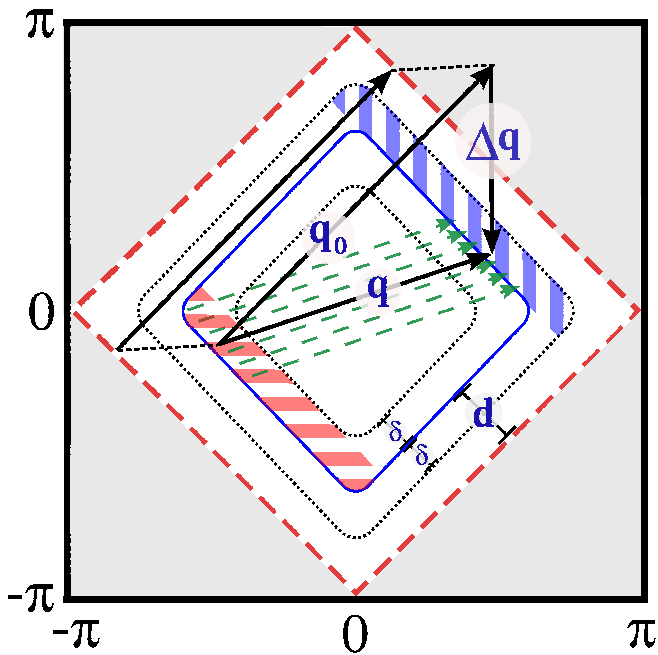}
\caption{
(Color online)
Illustration of the pairing model at small $U$ and $h$.
The half-filling FS is the large diamond (red dashed).
The non-interacting FS at low doping
remains approximately the shape of a diamond (blue solid).
AFM order arises from
$\mathbf{q}_0$, the pairing vector across the half-filling FS.
The pairing vector is
$\mathbf{q}$ across the doped FS.
The difference between $\mathbf{q}_0$ and $\mathbf{q}$,
$\Delta \mathbf{q}$,
determines the characteristic modulating wavelength of the l-SDW.
}
\label{2DFM}
\end{figure}

We first specify the pairing mechanism \cite{ZhangHF2008} more explicitly.
Recall that the non-interacting energy for the
state $\ket{\bfp}$ is
\begin{equation}
  \epsilon_\bfp = -2(\cos p_x+\cos p_y).
\label{Eksing}
\end{equation}
The plane-wave state is
$\ket{\bfp} = \frac{1}{\sqrt{N}} e^{i\bfp \cdot \bfr}$,
with $\mathbf{r}=(x,y)$, where $x$ and $y$ are integer coordinates denoting
lattice sites.
Consider a pair of spin-$\uparrow$ and
spin-$\downarrow$ electrons in a $\bfp$ state, where
$\epsilon_\bfp \le \epsilon_F$, with $\epsilon_F$ the
Fermi energy. In the pairing model \cite{ZhangHF2008}, this pair is made to
partially occupy a ${\bfp'}$ state:
\begin{eqnarray}
  \ket{\upa} &=& u_\bfp \ket{\bfp} + v_\bfp \ket{\bfp'}, \nonumber  \\
  \ket{\dna} &=& u_\bfp \ket{\bfp} - v_\bfp \ket{\bfp'},
\label{eq:pairingstate}
\end{eqnarray}
where $|u_\bfp|^2+|v_\bfp|^2=1$. This pair gives the following contribution to the local density in real space
\begin{eqnarray}
  n_\upa(\bfr,\bfp) &=& \frac{1}{N}\,( 1 + 2 |u_\bfp v_\bfp| \cos[ (\bfp'-\bfp) \cdot \bfr] ),\\
  n_\dna(\bfr,\bfp) &=& \frac{1}{N}\,( 1 - 2 |u_\bfp v_\bfp| \cos[ (\bfp'-\bfp) \cdot \bfr] ).
\label{eq:localdensity}
\end{eqnarray}
And an SDW state will result from the state in Eq.~(\ref{eq:pairingstate}),
with local spin
\begin{eqnarray}
  s(\bfr,\bfp) &\equiv& n_\upa(\bfr,\bfp) - n_\dna(\bfr,\bfp) \nonumber \\
  &=& \frac {4}{N}|u_\bfp v_\bfp|\cdot\cos[(\bfp'-\bfp) \cdot \bfr].
\label{eq:wave}
\end{eqnarray}
The SDW state lowers the interaction energy contribution of the pair
compared to the non-interacting solution (i.e., the solution when $v_\bfp=0$) by the amount:
\begin{eqnarray}
  \Delta\varepsilon_{\rm V}(\bfp) &=& U\sum_{\mathbf{r}}
  \left[ n_\upa(\bfr,\bfp) \, n_\dna(\bfr,\bfp) - \frac {1}{N}\frac{1}{N} \right] \nonumber \\
  &=& -\frac {U}{4} \sum_{\mathbf{r}} s^2(\bfr,\bfp).
\label{dEvsing}
\end{eqnarray}
If we have multiple pairs each formed as in Eq.~(\ref{eq:pairingstate}),
the change in interaction energy follows the same relation:
\begin{equation}
  \Delta E_{\rm V} = -\frac {U}{4} \sum_{\mathbf{r}} s^2({\mathbf r})
  =-\frac {U}{4} \sum_{\mathbf{r}}  \left[ \sum_\bfp s(\bfr,\bfp) \right]^2,
 \label{Ev_formula}
\end{equation}
where the sum over $\bfp$ is over all pairing plane waves (one of the four sides is illustrated by striped areas in
Fig.~\ref{2DFM}).

At half-filling,
the shell at the Fermi level, i.e. on the border of the diamond,
is open, with the number of degenerate $\bfp$ states
equal to twice the number of spin-$\uparrow$ (or spin-$\downarrow$)
electrons that need to be accommodated.
Pairing can be
achieved by choosing $\bfp'-\bfp=\mathbf{q}_0=(\pi, \pi)$, i.e.
having electrons occupy two states in the open shell across the FS.
This is perfect nesting and the SDW formed has perfect AFM order.
Because pairing occurs in the open
shell at the FS, the reduction in interaction energy from the SDW has
no penalty, i.e. no increase in the kinetic energy.

\subsection{The linear spin-density wave state}
\label{ssec:l-sdw-pairing}

We next consider
the case of low $U$, slightly doped ($U\ll$ bandwidth),
to help understand the mechanism of the l-SDW state.
As the FS shrinks with small doping,
we assume that it remains approximately the shape of a diamond.
The distance between the FS at half-filling and the doped FS
is determined by
\begin{equation}
  d = \frac {\sqrt{2}}{4}h\pi.
\label{distance}
\end{equation}

As the interaction is turned on, it can become advantageous
for some of the electrons near the FS to be partially excited.
Partially occupied states around the FS can then allow
pairing
across the FS,
which causes a correlation between electrons of opposite spins to generate
an SDW. The presence of the SDW
will lower the interaction energy.
However, in this more general case there will also be an increase in the kinetic
energy.
When the lowering of the interaction energy surpasses the increase in kinetic energy,
an overall lower energy state
is found compared to the free-electron (or RHF) solution.

We first determine the kinetic energy change.
At low $U$, pairing occurs near the FS. Electrons from a small region
immediately inside the FS are excited.
As a crude model \cite{Overhauser,ZhangHF2008},
we assume that a fraction $f$ of the electrons within a
distance $\delta$ of the
FS are excited, as illustrated by the horizontally striped region in Fig.~\ref{2DFM}.
The excited electrons occupy the region (vertically striped) immediately above the FS, also of
thickness $\sim \delta$.
We take $u_\bfp$ and $v_\bfp$ in the pairing state in  Eq.~(\ref{eq:pairingstate})
to be independent of $\bfp$:
$u_\bfp=u$ and $v_\bfp=v$. Thus
the vertically striped area has uniform density, and  $f=|v|^2$ .
An upper bound
to the kinetic energy
increase due to this process is easily estimated. It is, for each excited
electron, given by:
\begin{equation}
  \Delta\varepsilon_{\rm K}(\bfp) = \nabla_\bfp \epsilon_\bfp \cdot \Delta \bfp,
\label{dEksing}
\end{equation}
where $\Delta \bfp = \delta\:(1,1)\sqrt{2}/2$.
The total kinetic energy increase is then
\begin{eqnarray}
  \Delta E_{\rm K} &=& 8\,\,f\,\bar n_\sigma \int \Delta\varepsilon_{\rm K}(\bfp) \, dS \nonumber \\
             &=& \frac {N\delta^2}{\pi^2}\,8\,f\left[1+\cos \displaystyle\frac{h\pi}{2} \right],
\label{kinetic}
\end{eqnarray}
where $S$ in the integral is over the horizontally striped area inside the FS,
and the  factor of $8$ accounts for the $4$ sides and $2$ spin species.

We now determine the interaction energy change, and show that
the optimal SDW is along the $y$- or $x$-direction.
From Eq.~(\ref{Ev_formula}) we see that the maximum reduction is achieved by maximizing
 the quantity
 \begin{equation}
 I_{\rm V} = \sum_{\mathbf{r}} \left[\sum_{\mathbf{q}} \cos(\mathbf{q}\cdot\mathbf{r}) \right]^2,
\label{Eq:I_V_int}
\end{equation}
where the sum over $\mathbf{q}$ is over all pairing states, with
$\mathbf{q}=\bfp'-\bfp$.
This is realized if all the electron pairs line up their pairing vectors.
There are two groups of pairing states, corresponding to the two diagonal directions.
Within each group, the optimal choice is for all pairs to have one common pairing
wavevector $\mathbf{q}$. Let us denote the pairing wavevectors along $[11]$ and $[-11]$ by
$\mathbf{q}$ and  $\mathbf{q'}$, respectively, and write:
$\mathbf{q}=(\pi,\pi)-\Delta \mathbf{q}$ and $\mathbf{q'}= (-\pi,\pi)-\Delta \mathbf{q'}$.
We then obtain:
\begin{eqnarray}
I_{\rm V} &\propto& \sum_{\mathbf{r}}[\cos(\mathbf{q}\cdot\mathbf{r})
+\cos(\mathbf{q'}\cdot\mathbf{r})]^2 \nonumber  \\
&=& N+\sum_{\mathbf{r}}
(\cos[(\Delta\mathbf{q}+\Delta\mathbf{q'})\cdot\mathbf{r}] \nonumber \\
&&+\cos[(\Delta\mathbf{q}-\Delta\mathbf{q'})\cdot\mathbf{r}]).
\label{eq:pairingItf}
\end{eqnarray}
The maximum is achieved in Eq.~(\ref{eq:pairingItf}) when $\Delta\mathbf{q}
=\pm \Delta\mathbf{q'}$. This occurs when $\mathbf{q}$
and $\mathbf{q'}$ are such that the SDW modulation from the two groups of pairing states
are the same, leading to a positive `interference' between them.
The direction of the modulating wavevector must be along $[01]$ (or $[10]$).
The magnitude is given by
\begin{equation}
|\Delta\mathbf{q}|=2\sqrt{2}\,d=h\pi,
\label{eq:Deltaq}
\end{equation}
as illustrated in Fig.~\ref{2DFM}.
This leads to the following total reduction in interaction energy:
\begin{equation}
  \Delta E_{\rm V}
             = -\frac {N\delta^2}{\pi^2}4 |u|^2\,fU\left(1-\frac{h}{2}\right)^2.
\label{potential}
\end{equation}
Thus the lowest energy state is an
l-SDW with broken $x$-$y$ symmetry, with the modulation
 along either the $x$- or the $y$-direction. The modulating wavelength
is $\lambda_{\rm l-SDW}=2/h$, consistent with our numerical result.

To reach an SDW state of lower energy than the non-interacting solution, the condition
\begin{equation}
  |\Delta E_{\rm V}| \geq |\Delta E_{\rm K}|
\label{cond}
\end{equation}
must be satisfied. From Eqs.~(\ref{kinetic}) and (\ref{potential}), we
obtain
\begin{equation}
  |u|^2U\left(1-\frac{h}{2}\right)^2 \geq 2\left[1+\cos \displaystyle\frac{h\pi}{2}\right].
\label{condU}
\end{equation}
Taking $|u| \sim 1 $ on the left-hand side,
we obtain a rough estimate to the critical value which $U$ must exceed:
\begin{equation}
  U_{\rm c} = \frac{2\left[1+\cos \displaystyle\frac{h\pi}{2} \right]}{\left(1-\displaystyle\frac{h}{2}\right)^2}.
\label{Ucri}
\end{equation}

\begin{figure}
\includegraphics[width=\columnwidth]{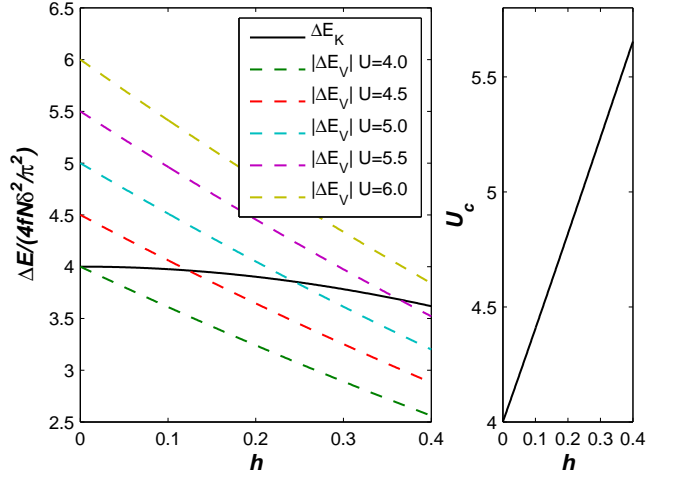}
\caption{
(Color online)
Absolute values of kinetic energy gained
and interaction energy lost in the pairing model.
On the left the energies are plotted
as a function of $h$ for several values of $U$.
On the right, $U_{\rm c}$ is plotted vs. $h$.
}
\label{UcHc}
\end{figure}

The absolute value of the kinetic and interaction energy
changes in Eqs.~(\ref{kinetic}) and (\ref{potential}) are plotted
vs. $h$ in the left
panel in Fig.~\ref{UcHc}.
$\Delta E_{\rm K}$ is independent of $U$, while $\Delta E_{\rm V}$ is proportional to $U$,
for which several curves are plotted for various values of $U$.
It is seen that a critical value of $U$ exists for doped system ($h\ne 0$).
Above $U_{\rm c}$, the two curves cross
at a critical $h_{\rm c}$, below which the broken-symmetry l-SDW state exists.
As $U$ increases, the point of crossing, $h_{\rm c}$, moves to the right.
Equivalently, the critical $U_{\rm c}$ decreases as doping is reduced.
In the right panel the curve of $U_{\rm c}$ vs. $h$ is plotted to illustrate this.

\subsection{Comparison with numerical results}
\label{ssec:l-sdw-compare-model-v-data}

\begin{figure}
\includegraphics[width=\columnwidth]{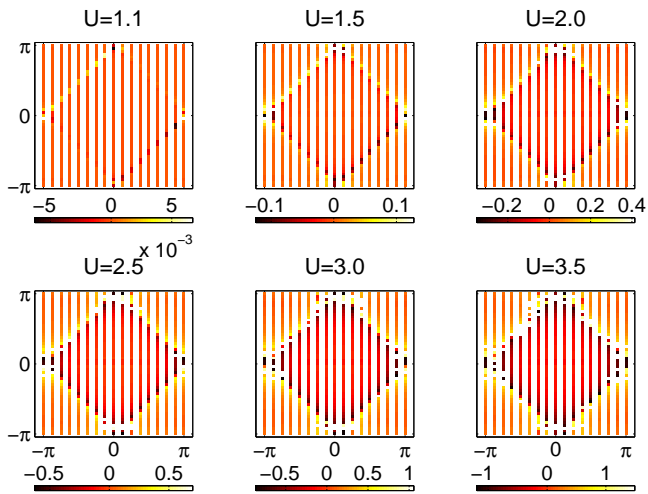}
\caption{
(Color online)
Energy plot of the modification to the RHF band structure in the UHF solution, for a sequence
of $U$ values in the l-SDW regime. Shown are the values
$\lambda_{\sigma\bfp}-\epsilon_\bfp$ vs. ${\mathbf p}$, where $\epsilon_\bfp$
is given in Eq.~(\ref{Eksing}).
The system is a $16\times 48$ supercell with doping of $h=1/24$.
}
\label{fig:lambda_k-diff}
\end{figure}

The simple model and analysis above
capture most of the properties of the exact UHF ground
state at low $U$ and small $h$. It gives the correct l-SDW modulating
wavelength, and explains the existence of $U_{\rm c}$ and how it varies with
doping. Because of the crudeness of the model, the values of $U_{\rm c}$
and other quantitative features are not very accurate compared to the
exact numerical results in Sec.~\ref{ssec:l-sdw}. Larger discrepancies can be expected further
away from its domain of validity, namely small
doping and modest interaction (although it also incorrectly predicts $U_{\rm c}=4$
as $h\rightarrow 0$).

The model considers only
pairing of two electrons, so CDW is excluded. This is consistent
with the numerical result that at low $U$, CDW is much weaker than SDW order.
The exact UHF solution will necessarily involve
more electrons in the pairing \cite{ZhangHF2008}, which will
lead to a larger energy lowering $|\Delta E_{\rm V}|$ (and thus lower $U_{\rm c}$)
and will result in CDW, as observed in the numerical results.

Figure~\ref{fig:lambda_k-diff} shows the modification to the RHF band structure in the
UHF solution as a function of interaction strength. The difference between the UHF eigenvalue
$\lambda_{\sigma \bfp}$ and the RHF spectrum $\epsilon_\bfp$ in Eq.~(\ref{Eksing}) is plotted for all momentum
values ${\mathbf p}$. As discussed in
Fig.~\ref{HF_bandstructure}, the
eigenvalue  $\lambda_{\sigma \bfp}$ is identified with the momentum ${\mathbf p}$ with which
the corresponding eigenstate has the maximum overlap. We see that, just above $U_{\rm c}$,
a small fraction of the states on the FS are involved in pairing, which creates a small
energy lowering that leads to the UHF solution. The plot is for a single twist angle. In a finite
system, the shift in momentum space from the twist creates a small asymmetry between each pair of
surfaces diagonally across. At small $U>U_{\rm c}$, this is reflected in the solution as an asymmetry
in the gaps on the two surfaces. As $U$ increases, excitation spans a wider region at the FS, and the gap structure from pairing becomes more pronounced.

\begin{figure}
\includegraphics[width=\columnwidth]{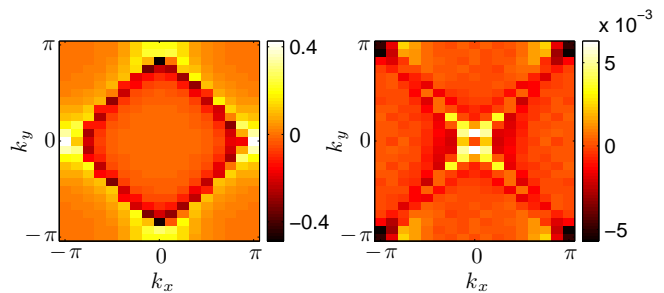}
\caption{
(Color online) Density plots of $\Delta n({\mathbf p})$, the momentum distribution difference
from RHF solution (left) and its correlation $\Delta n({\mathbf p})\Delta n({\mathbf p'})$ (right).
The system is a $16 \times 24$ suppercell
with doping of $1/12$ at $U=3.0$.
Negative peaks at $(\pm\pi, \pm(\pi-\pi/12))$ in the correlation
result from the pairing.
}
\label{kspace}
\end{figure}

The momentum distribution from a numerical UHF solution at $h=1/12$ is shown
in Fig.~\ref{kspace}.
The left panel plots $n({\mathbf p})$ minus
the non-interacting value $n_0({\mathbf p})$:  $\Delta n({\mathbf p})=n({\mathbf p})-n_0({\mathbf p})$.
Electrons are excited from the darker area to the lighter.
The right panel shows the two-point correlation function from the left panel:
$\Delta n({\mathbf p})\Delta n({\mathbf p'})$ vs. $(\bfp-\bfp')$.
Negative peaks are seen at $(\pm\pi, \pm(\pi-\pi/12))$ on the right, which
result from the pairing between the negative
just inside the FS
(where electrons are excited from)
and the positive immediately above the FS
(where electrons are excited to) in the left panel.
The position of the negative peaks
indicates a pairing vector of $\Delta \mathbf{q}=(0, h\pi)$,
consistent with the pairing vector in the analytical model.

\subsection{Diagonal spin-density wave states}
\label{ssec:D-sdw-model}

As mentioned before, diagonal modulations lose
the interference between $[11]$ and $[-11]$, so a
diagonal (or any orientation other than $[10]$ and $[01]$)
SDW is not the solution at small $h$ and moderate $U$.
This does not exclude it as a solution as we move away from this parameter regime,
when the distortion to the FS becomes more severe.

\begin{figure}
\includegraphics[width=\columnwidth]{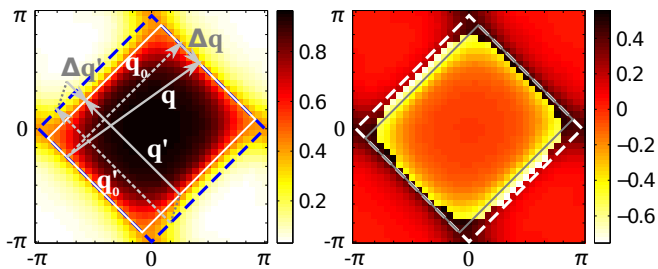}
\caption{
(Color online)
Illustration of the pairing scheme for d-SDW order.
The left panel shows $n({\mathbf p})$ and the right panel $\Delta n({\mathbf p})$,
the  difference from the non-interacting solution. The momentum distribution is actual
numerical data from a system of $36\times36$ with doping of $1/6$ at $U=5.0$.
Electrons are excited from the FS across the $(-\pi, \pi)$-direction
to the FS across the other direction $(\pi, \pi)$, such that the FS along the latter reaches
the half-filling FS.
This enables two groups of pairings to maintain interference,
with $\Delta \mathbf{q}=\Delta \mathbf{q'}$, to lower the energy.
}
\label{2DdiagFM}
\end{figure}

This situation happens when the doped FS is deformed sufficiently
away from the half-filling shape of a
diamond and the area of excitation becomes sufficiently
large to reach the half-filling FS.
The number of pairs that could participate in the
 `interference'
  of the l-SDW is decreased, because the FS no longer has the shape of a diamond.
 Eventually it becomes energetically more favorable to have
the FS be longer in one diagonal direction than the other, i.e., to break the four-fold rotational symmetry.
As illustrated in Fig.~\ref{2DdiagFM},  it is then possible to create two
different types of pairing states along the two diagonal directions, such that they share
a common modulating wavevector along one diagonal
direction:  $\Delta \mathbf{q} = \Delta \mathbf{q'}$.
The two groups of pairs will achieve interference, similar to the case of l-SDW.
As in Sec.~\ref{sec:model}, the pairing vector is determined by $h$,
giving $\Delta \mathbf{q} =(h\pi, h\pi)$.
which gives rise to an SDW with modulating wave along $[11]$-direction,
and of wavelength $\lambda_{\rm d-SDW}=\sqrt{2}/h$.
The corresponding wavelength for d-CDW is $1/\sqrt{2}h$.
This is consistent with the numerical results in Sec.~\ref{ssec:d-sdw-stripe}.

\section{Discussion}
\label{sec:discussion}

\begin{figure}
\includegraphics[width=\columnwidth]{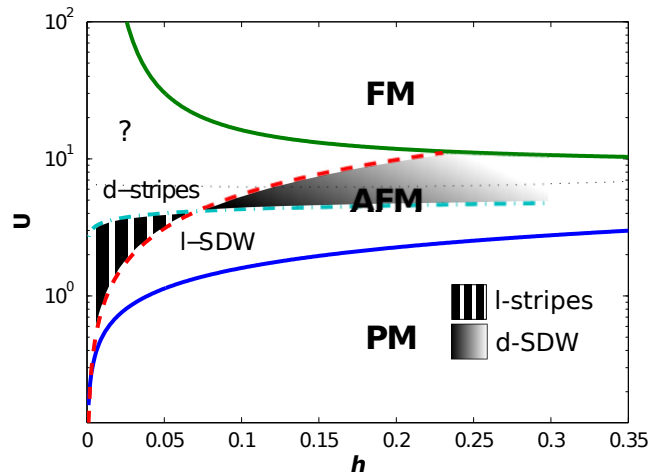}
\caption{
(Color online)
Phase diagram of the ground state of the 2D Hubbard model from
UHF.
The phase boundaries are determined by fitting our numerical results,
and are meant only as rough guidelines.
Solid lines separate the antiferromagnetic (AFM)
phase from the paramagnetic (PM) phase and
the ferromagnetic (FM) phase.
Within the AFM phase, the different regions include: l-SDW (SDW state with
a linear modulation along $[10]$-direction);
l-stripes (density saturation to $1$, with linear modulation along the $[10]$-direction);
d-SDW (SDW state with a modulating along the diagonal $[11]$-direction);
d-stripes (density saturation to $1$, with diagonal modulation).
The black dotted line gives the theoretical estimate (Stoner criterion) for the
transition from the RHF solution (PM) to FM.
}
\label{2D_phasediag}
\end{figure}

We can now place our 2D results in the context of an HF phase diagram for the Hubbard model.
Our numerical calculations have
focused on small and intermediate dopings ($h$ from
$0$ to $\sim0.3$), and small to moderate interactions ($U$ from $0$ to $\sim10$),
because of possible connections with the many-body ground state at moderate interacting
strengths.
The analytic calculations are for small $h$ and low $U$, where
our pairing model captures the physics in the HF framework.
Our numerical results are sufficiently detailed such that we could determine
some phase boundaries as shown in Fig.~\ref{2D_phasediag}. 
We fitted the numerical locations for the phase transition or crossover 
using power functions, 
except for the AFM to FM transition which was fitted by an exponential.
Because of the limited number of data points and the finite resolution 
with which the transition was scanned, there are significant
uncertainties in the fits, of several line widths or larger. The phase
boundaries are thus only meant as rough guides.

At half-filling, the UHF solution
is an AFM state.
Upon doping, there is a phase boundary $U_{\rm c}(h)$,
shown as the blue line in Fig.~\ref{2D_phasediag},
below which is the PM phase.
Above $U_{\rm c}(h)$ is an AFM region where a rich set of sub-regions exhibit different
characters, including the l-SDW
states we have focused on in this work; we describe this region in further detail below.
Above the AFM phase is an FM phase. Our numerical UHF calculations
 show that the FM state has lower energy above the green solid line.
 The RHF approach, naturally, predicts an earlier
transition to FM. This is the theoretical phase boundary from Stoner criterion,
and is shown as the black dotted line.
Recall that we have excluded spiral SDWs.
As we discussed, this is not the ground state 
at low $U$ (see also Refs.~\onlinecite{Schulz1990, Ichimura1992}).
However, at large $U$, spiral orders can become more favorable deep in the 
d-stripes region.

Between the PM and FM phases is the AFM phase.
In this region, at low and intermediate $U$, we see
an l-SDW state with a long wavelength modulation along the $[10]$-direction;
a weaker CDW accompanies the SDW.
Near half-filling, as $U$ is increased the l-SDW state evolves into
a l-stripes state which shares the same characteristic wavevector as the l-SDW, but whose
CD saturates to $1$ in regions separated by `stripes' anchored by the nodal positions defined by the SDW. The holes are localized in these stripes.
This is consistent with the observation in Ref.~\onlinecite{Littlewood1991} of SDW
deforming into domain walls with increasing $U$.
The transition from delocalized holes
(such as the l-SDW state) to localized holes is denoted by the red dashed line in
Fig.~\ref{2D_phasediag}.
As we move further away from half-filling, the l-SDW at lower interaction
changes its direction of modulation as $U$ is increased. This forms a d-SDW state.
The transition from a state with
 modulation along the $[10]$-direction
 to one with diagonal modulation
is denoted by the cyan
 dot-dashed line. We see that the two dash lines cross each other. At low doping
 ($h \lesssim 0.1$), the
 system reaches an l-stripes state first before changing the direction of modulation
 to a d-stripes state.
At higher doping, the order is reversed. The l-SDW first changes into a d-SDW state. As $U$ is
further increased, density saturation appears, and holes become localized in a d-stripes state.

It is important to keep in mind that the results we have discussed and the phase diagram
above are for HF theory. For strong interactions in particular, the HF results are expected to
be severely biased and correlation effects can fundamentally change the nature of the
many-body state. For example, the FM phase was shown not to exist at low density
($h>0.5$) in the 3D Hubbard model \cite{ChangFM_PRA2010}.

The present work was in part
motivated by a recent quantum Monte Carlo (QMC)
calculation \cite{Chang2010} which indicated that
the ground state of
the 2D Hubbard model has a long wavelength SDW collective mode.
Upon doping, the AFM order at half-filling was found to evolve into
an SDW state with a long wavelength modulation which has essentially
a constant charge-charge correlation at low to intermediate interacting strengths.
Given that the UHF solution is qualitatively correct at half-filling,
it was natural to ask to what extent
the UHF solution contains any of these features upon doping.

We see from the numerical results in this work that the UHF solution appears
to qualitatively capture the basic features of the magnetic
correlations in the
ground state upon doping, as it does at half-filling.
Of course the UHF solution gives a static modulated SDW,
while the many-body ground state in the QMC
preserves translational invariance and the SDW correlation is only
seen in the correlation functions \cite{Chang2010}.
This is similar to the situation at half-filling.

In the UHF solution, the tendency for the holes to localize is much
overestimated. This was part of the reason to focus on low $U$ in
the present study.
A CDW correlation almost always accompanies the SDW in the UHF solution,
and holes appear to localize (leading to domain walls or stripes)
at $U\sim 4$. In contrast,
holes remain delocalized (wave-like) in the
many-body solution \cite{Chang2010},
with essentially constant charge-charge correlation, until
the strong interaction regime ($U\gtrsim 10$).
It is an interesting question whether diagonal order, which is present in the HF solution
at larger $U$, is  present in the true many-body ground state.

The UHF solution thus provides a useful starting point for understanding
the magnetic and charge correlations in the ground state of the
Hubbard model at intermediate interactions. In addition, the ability
to reliably determine the true UHF ground state numerically could
prove valuable in QMC calculations, which often require
a trial wavefunction and where the qualititative correctness of the
trial wavefunction can make a significant difference.
Although the physics in the UHF solution is sensitive
to the particular many-body Hamiltonian, the basic approach we have used and
the basic ideas of the analytic calculations are general
(see also Ref.~\onlinecite{ZhangHF2008})
and can be expected to find applications in other many-fermion systems.

\section{Conclusion}
\label{sec:conclusion}

In summary, we have performed exact numerical calculations for the
UHF ground state of the Hubbard model systematically for a wide range of lattice sizes,
initial conditions, doping and interaction strengths. 
Special care has been taken to
reduce finite-size effects in order to obtain the solution at the
thermodynamic limit. These results
allow us to map out the magnetic phase diagram for regimes most relevant in
modeling condensed matter systems.

A broken-symmetry UHF solution exists above a
critical $U_{\rm c}$, whose value increases with doping.  Above $U_{\rm c}(h)$,
the ground state is a static l-SDW/CDW, with a modulation whose
wavelength is inversely proportional to doping at small $h$.  The
amplitude of the SDW/CDW decreases with $h$ and increases with $U$. At
low $U$, the SDW amplitude is much stronger than that of the
accompanying CDW, and the holes are essentially delocalized. For
larger $U$, the SDW and CDW amplitudes become more comparable.  At
small doping, the solution turns into the l-stripes state with the
same characteristic modulating wavevector and holes localized at the
nodal positions, before eventually entering the d-stripes state.  At
larger doping, the l-SDW state first turns into the d-SDW state before
eventually entering the d-stripes state at larger interactions.

We have also presented an analytic theory to explain the mechanism for
the formation of the SDW state. The model provides a conceptual
understanding of the physics of SDW which can be applied in systems 
beyond the 2D Hubbard model. Comparison with
recent QMC results shows that the UHF solution captures the magnetic 
correlations in the true many-body ground state at intermediate interactions.

\begin{acknowledgments}
The work was supported in part by NSF
(DMR-0535592 and DMR-1006217)
and ARO (56693-PH).
Computational support provided by the Center for Piezoelectrics by Design.
We thank H.~Krakauer,  B.~Normand, and E.~Rossi
for useful discussions.
S.Z., J.X., and C.C. are grateful to Professor T.~Xiang and the Institute
of Physics, Chinese Academy of Sciences, and to Professor X.Q.~Wang and
Renmin University of China for hospitality during an extended visit,
where part of the
work was performed.
\end{acknowledgments}

\end{document}